\documentclass[preprint,showpacs,preprintnumbers,amsmath,amssymb]{revtex4-1}
\usepackage{graphicx}
\usepackage{dcolumn}
\usepackage{subfigure}
\usepackage{nicefrac}
\usepackage{bm}
\usepackage[normalem]{ulem}	
\usepackage{natbib}      
\usepackage{array,multirow,amsmath,amsfonts}
\usepackage{color}

\newcolumntype{C}[1]{>{\centering\let\newline\\\arraybackslash\hspace{0pt}}m{#1}}

\def\4kBT{$4k_\textrm{B}T$}
\def\In5{In$_5$}
\def\B6{B$_6$}

\definecolor{dkgreen}{rgb}{0.21,0.49,0.16}

\begin{document}
\preprint{}
\title{Direct visualization of coexisting channels of interaction in CeSb}

\author{Sooyoung Jang}
\email[]{Email address: sjang@lbl.gov}
\affiliation{Advanced Light Source, Lawrence Berkeley National Laboratory, Berkeley, CA 94720, USA, Department of Physics, University of California, Berkeley, California 94720, USA, Materials Sciences Division, Lawrence Berkeley National Laboratory, Berkeley, California 94720, USA}

\author{Robert Kealhofer}
\author{Caolan John}
\author{Spencer Doyle}
\affiliation{Department of Physics, University of California, Berkeley, California 94720, Materials Sciences Division, Lawrence Berkeley National Laboratory, Berkeley, California 94720, USA}

\author {Jisook Hong}
\affiliation{Department of Chemistry, POSTECH, Pohang 37673, Korea}  

\author {Ji Hoon Shim}
\affiliation{Department of Chemistry and Division of Advanced Nuclear Engineering, POSTECH, Pohang 37673, Korea}  

\author {Qimiao Si}
\affiliation{Department of Physics and Astronomy, Rice University, Houston, TX 77005, USA}  

\author {Onur Erten}
\affiliation{Max Planck Institute for the Physics of Complex Systems, D-01187 Dresden, Germany}  

\author{J. D. Denlinger}
\email[]{Email address: jddenlinger@lbl.gov}
\affiliation{Advanced Light Source, Lawrence Berkeley National Laboratory, Berkeley, CA 94720, USA}

\author{James. G. Analytis}
\email[]{Email address: analytis@berkeley.edu}
\affiliation{Department of Physics, University of California, Berkeley, California 94720, USA, Materials Sciences Division, Lawrence Berkeley National Laboratory, Berkeley, California 94720, USA}

\noindent

\vspace{10 pt}
\vspace{10 pt}

	
\begin{abstract}

 \textbf{
Our understanding of correlated electron systems is vexed by the complexity of their interactions. Heavy fermion compounds are archetypal examples of this physics, leading to exotic properties that weave together magnetism, superconductivity and strange metal behavior. The Kondo semimetal CeSb is an unusual example where different channels of interaction not only coexist, but their physical signatures are coincident, leading to decades of debate about the microscopic picture describing the interactions between the $f$ moments and the itinerant electron sea. Using angle-resolved photoemission spectroscopy, we resonantly enhance the response of the Ce$f$-electrons across the magnetic transitions of CeSb and find there are two distinct modes of interaction that are simultaneously active, but on different kinds of carriers. This study is a direct visualization of how correlated systems can reconcile the coexistence of different modes on interaction - by separating their action in momentum space, they allow their coexistence in real space.}

\end{abstract}

\maketitle

{\bf Introduction.} One of the earliest triumphs of the theory of many-body systems was the Kondo model of dilute magnetic impurities embedded in a metal~\cite{kondo64}. The theory describes the scattering of the conduction electrons by local moments via Kondo exchange, causing an increase in the resistance until, at some low temperature $T_K$, the conduction electrons become entangled with these moments, forming a neutral singlet that no longer scatter the carriers. Doniach suggested that this idea could be extended to some classes of lanthanide and actinide compounds that form lattices of local $f$-moments, in which case the singlets themselves should conserve crystal momentum and form a heavy fermi liquid at the Kondo coherence temperature~\cite{doniach77_1,mott74, doniach77}. However, the Rudderman-Kittel-Kasuya-Yoshida (RKKY) interaction, which is closely related to Kondo exchange, can also favor the magnetic ordering of the magnetic ions, and so the $f$-electrons must choose between these two paths. In principle, magnetic ordering and Kondo entanglement should be mutually exclusive, but in general signatures of both mechanisms are evident in most materials, as if the two states coexist. The semi-metal CeSb is a case where these signatures appear to be coincident, raising the question of how the $f$-moments decide on the most favorable channel of interaction ~\cite{coqblin_exchange_1969,cooper_contrasting_1982,kioussis88}. The debate turns on understanding the complex interplay between symmetry and exchange, highlighting the importance of symmetry sensitive measurements of the interacting $f$-electrons.

\begin{figure*}[t]
\begin{center}
\includegraphics[width=18cm]{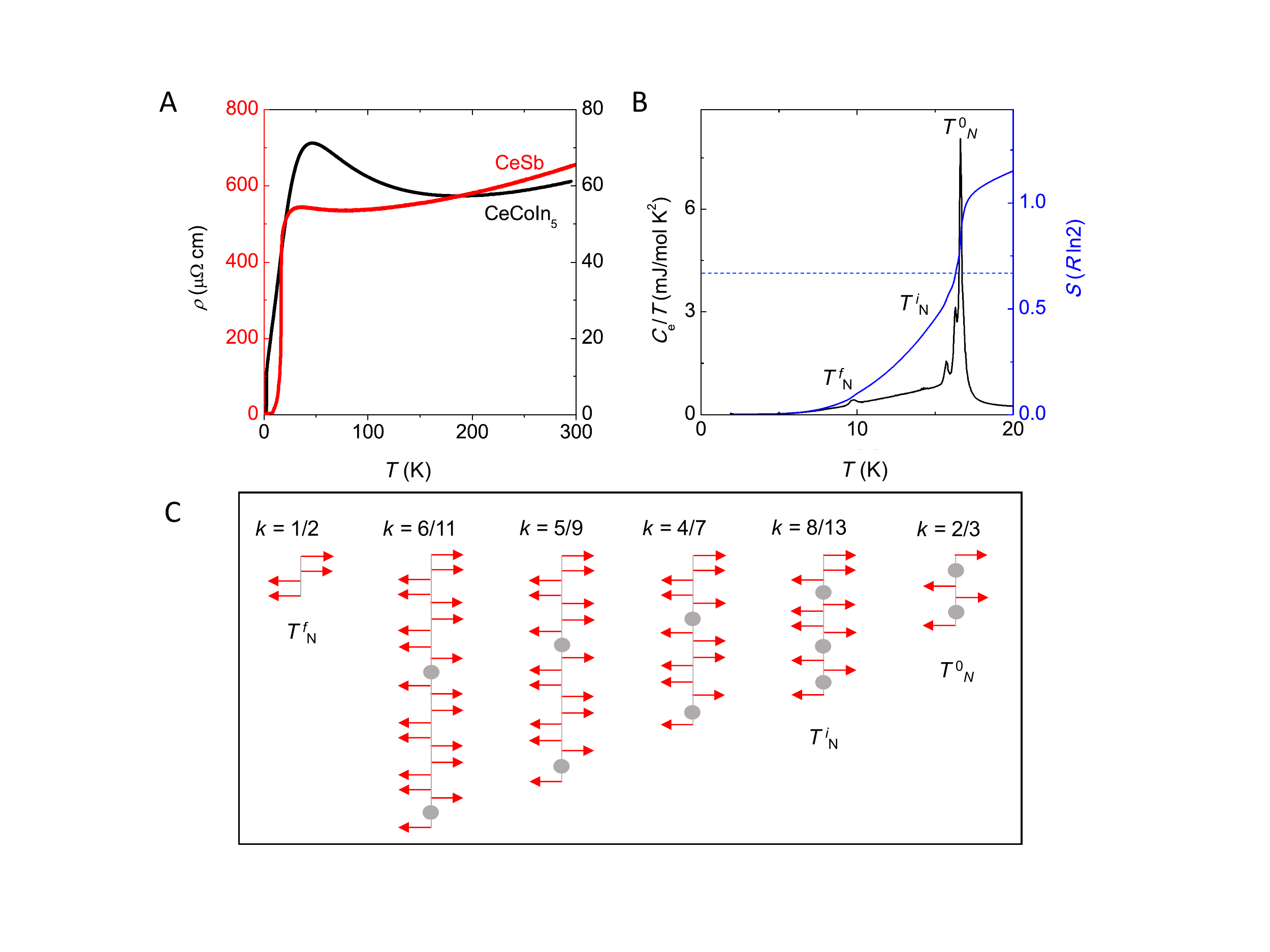}
\caption{{\bf Kondo semimetals in CeSb} {\bf A} Electronic specific heat ($C_e$) (black line with left axis) shows several magnetic transitions, from the antiferro-paramagnetic transition (AFP) at $T^0_N$ = 17 K to the  antiferro-ferrmagnetic transition (AFF) phase transitions at $T^f_N$ = 8 K. Entropy(S) (blue line with right axis) has been estimated by integrating the electronic specific heat leading to 15 K of the Kondo temperature ($T_K$). {\bf B} Temperature dependence of the electrical resistivity for semimetal CeSb and typical Kondo system CeCoIn$_5$, coherence peak at $T^*$ $\sim$ 35 K and 45 K signalling the presence of the Kondo scattering at high temperature for both compounds, respectively. However, the residual resistivity for CeSb is a lot enhanced than that of CeCoIn$_5$ due to its semietallic nature. {\bf C} Schematic of magnetic structure for each transitions. Red arrows and grey circles present the direction of each ferromagnetic- and paramagnetic-layers, respectively.}
\label{fig:fig1}
\end{center}
\end{figure*}

 The high temperature paramagnetic state of CeSb is characterized by an enhanced heat capacity with a Sommerfeld coefficient $\gamma\sim 450$ mJ/molK$^2$ and a huge resistivity (with residual $\sim 600 \mu\Omega$cm), as shown in Fig. \ref{fig:fig1}A and B. As the temperature decreases there is slight upturn of the resistivity, consistent with Kondo scattering~\cite{guo_possible_2017}, preceding a precipitous drop in both the heat capacity and the resistivity of the system. This drop begins well above the magnetic transition occurring at $T_N^0$, which is the first of a cascade of $i$ transitions with critical temperatures $T_N^i$. Each of these magnetic phases is composed of ferromagnetic (FM) and paramagnetic (PM) layers of Ce moments, stacked anti-ferromagnetically in the [001] crystallographic direction with a well defined wavevector $k$. Each transition can be thought of a precursor of the lowest temperature phase at $T_N^f$ illustrated in Fig.~\ref{fig:fig1}C, known as type-IA AFM~\cite{Rossat77}. As temperature increases, the $i$th transition will gain a PM layer, reducing the translational symmetry. The onset of magnetic order only accelerates the suppression of the heat capacity and resistivity, settling on a low temperature Sommerfeld coefficient of $\sim $20 mJ/molK$^2$ and residual resistivity $\rho_0\sim 2~\mu\Omega$cm, both more than two orders of magnitude smaller than the values extrapolated from the high temperature state.  

\begin{figure*}[t]
\begin{center}
\includegraphics[width=16cm]{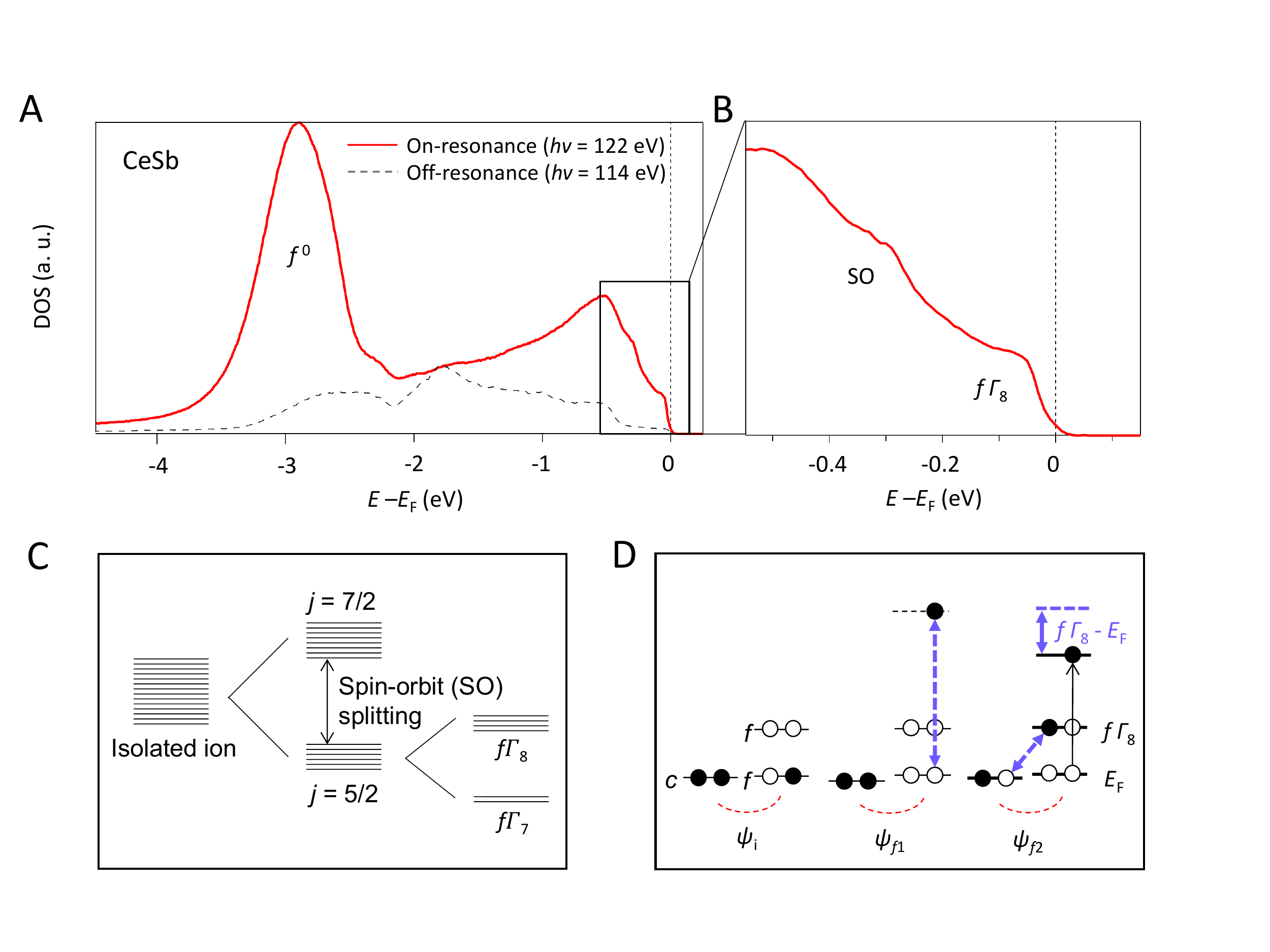}
\caption{{\bf Crystal electric field splitting} {\bf A-B} The $k$-integrated $f$-density of state (DOS) for on- and off-resonance ARPES data are displayed by red solid line and grey dashed line, respectively. Each $f$-state are indicated by $f\Gamma_8$ crystal electric field (CEF), spin-orbit (SO) side band peak, and the $f^0$ final state peak in the range of binding energy ($E_B$ = $E$ - $E_F$) between 0.1 and -4 eV. {\bf C} Crystal field scheme of cubic CeSb. SO separates $J$ = \nicefrac{7}{2} and \nicefrac{5}{2} multiplets and the latter is split by CEF into doublet $f\Gamma_7$  and quartet $f\Gamma_8$ manifolds, with $f\Gamma_7$ forming the ground state. {\bf D} Schematic of final state shakeup transitions for Ce 4$f^1$ including initial state hopping between the conduction electrons ($c$) and the $f$-states. Excitation into $f\Gamma_8$ CEF (or $J$ = 7/2 SO-split) $f$-states results in lowered photoelectron kinetic energies at the detector and the appearance of these $f$ states below $E_F$. 
}
\label{fig:fig2}
\end{center}
\end{figure*}

{\bf The interplay of symmetry and exchange in CeSb.} The signatures of magnetic order and Kondo-like behavior are essentially coincident; as the temperature is lowered, the first magnetic transition occurs at $T_N^0=16.5$ K, which is preceded by a downturn in the resistivity and the release of $\sim$ 0.65$R$ln2 magnetic entropy at $T_{\tilde{K}}\sim16~K$. The latter effects are the signatures that resemble the properties of Kondo metals but because the Kondo interactions must be weak in CeSb, we use the notation $\tilde{K}$ to distinguish it from the Kondo coherence that appears in other systems. To qualify this, we compare the resistivity of CeSb to that of CeCoIn$_5$, an archetypical Kondo metal, which shows a much more dramatic Kondo upturn in the resistivity. Note however that absolute values of the resistivities differ by one order of magnitude, which is reasonable given the carrier density is about 100 times smaller in CeSb, while its effective mass is only 10 times smaller~\cite{chechelsky17, hundley04, canfield00}. The Kondo behavior in CeSb is weak because the itinerant carriers are few, consisting of small electron-like and hole-like Fermi surfaces. Since the Fermi momentum $k_F$ is therefore small, FM correlations are favored by the RKKY interaction~\cite{ueda90}, consistent with the in-plane FM order. The unusual nature of the AFM transitions is more difficult to understand, but it is thought to arise out of the the interplay of symmetry and exchange in the ground state interactions. The $f$-ground state has $\Gamma_7$ type symmetry~\cite{heer_neutron_1979};  as shown in Fig. \ref{fig:fig2}C, spin-orbit (SO) coupling separates the $f$ states into \nicefrac{7}{2} and \nicefrac{5}{2} multiplets, and the latter is split by the crystal electric field (CEF) into a $\Gamma_8$ quartet and a $\Gamma_7$  doublet. The hole-like Fermi pockets from the Sb-band also have $\Gamma_8$ symmetry, which leads to a large interaction with the quartet. Kasuya {\it et al.} have argued that while Kondo exchange is important to understand the overall properties of CeSb, a critical mechanism behind the magnetic anisotropy of the ordered state is $p-f$ hybridization, which pulls down the $f\Gamma_8$ states so that they mix with the ground state ~\cite{kasaya83, Takahashi85}. Other authors have argued for non-Kondo exchange, mediated by both electrons and holes and independent of their symmetry~\cite{coqblin_exchange_1969, cooper_contrasting_1982}, though $p-f$ hybridization is still invoked to explain the appearance of PM layers ~\cite{kioussis88}. These latter proposals do not require an active Kondo interaction, leaving the observed Kondo signatures to be explained by some alternative mechanism. Both pictures require that the CEF is small, and indeed this has been measured to be $\sim 3$ meV~\cite{heer_neutron_1979}. Distinguishing these pictures and explaining the duplicitous properties of CeSb, requires symmetry-resolved measurements of the interaction between the Ce $f$-moments and the hole and electron Fermi pockets. To this end, we turn to angle-resolved photoemission spectroscopy(ARPES).

{\bf Angle-Resolved Photoemission Spectroscopy (ARPES).} Recent ARPES studies on CeSb have largely focused on verifying whether the system contains a Weyl-like crossing at the $X$-point of the Brillouin zone (BZ), ostensibly arising from a $p-d$ band inversion~\cite{hasan16, kaminski17, sato17}, but only a few have addressed the question of the magnetism itself~\cite{takahashi09} (we comment on the possible presence of Weyl-like features in the Supplementary Information.) The three dimensional nature of the Fermi surface creates a number of complications in obtaining reliable data on CeSb; the region of the BZ probed depends strongly on the photon energy, bulk bands and surface projections of those bands must be carefully distinguished  and orbital selection rules couple some bands preferentially to a given photon polarization and energy (see Supplementary Information for more details). Therefore, we perform our measurements at different polarizations, linear horizontal light (LH) and linear vertical light (LV), and photon energies ranging between 30 eV to 158 eV to access the intrinsic bulk behavior. Importantly, we focus on using photons that couple resonantly to the Ce $4d\rightarrow 4f$ transition ($h\nu = $122 eV), which strongly enhances the contribution of the $f$-states. This can be broadly thought of as orbital-specific ARPES, because it enhances the features related to the Ce $f$-state interaction across the magnetic transitions.

Figure \ref{fig:fig2}A shows several prominent Ce 4$f$-related features in the angle-integrated density of states (DOS), which can be distinguished by their absence in the off-resonance data. There is a strong feature at binding energy $E_B$ $\sim$ 2.9 eV, corresponding to the $f^0$ final-state peak associated with the cost of removing one electron from the trivalent Ce ion (4$f^1$ $\rightarrow$ 4$f^0$). In addition, there are two other features at $E_B$ $\sim$ 300 meV and $E_B$ $\sim$ 50 meV. The former is consistent with the known SO splitting of the $f$ states in Ce systems, suggesting the latter corresponds to the CEF splitting of the \nicefrac{5}{2} sextet. These states are actually above $E_F$, but can be observed in photoemission due to `final state excitations', explained schematically in Fig. \ref{fig:fig2}D\cite{hufner95}. In essence, the energy of the electrons reaching the detector depends on the final state of the system after the electron is removed, so that multi-electron excitations to intermediate states lead to lower kinetic energy photoelectrons. The result is that SO and CEF excited states appear at binding energies below $E_F$, mirroring their excitation energies above $E_F$ (distinguished as $\Gamma_8^*$ states). This observation alone indicates that Kondo-like many-body effects are present by following argument; note Ce$^{3+}$ only has one electron $f^1$, so that the electrons excited to the intermediate states must therefore originate from the itinerant electrons or holes, even though the photon is resonant to $f$-states. Nevertheless, even though this is consistent with a Kondo-like entanglement of local moments with itinerant carriers, there is no clear Kondo resonance observed near the Fermi energy, suggesting the Kondo effect is highly suppressed. This suppression of the $E_F$ Kondo resonance causes a ``pseudo-gap-like" appearance in the $f$-DOS consistent with single-impurity spectral function predictions for the narrow width and deep binding energy of the $f^0$ peak as shown in the Supplementary Information.

The 50 meV feature is an order of magnitude larger than the CEF splitting reported in earlier studies of CeSb~\cite{Rossat77, Rossat85}. However, a $\sim$50 meV CEF field is both consistent with other Ce$^{3+}$ compounds~\cite{inosov16}, and with known trends in rare-earth pnictides~\cite{birgeneau_crystal_1973}. The early neutron measurements of CeSb did not explore such large energy scales, and the 3 meV feature, which appears as a shoulder on a large elastic peak, may correspond to an additional splitting. Our results suggest that neutron studies should be revisited to explore higher energy scales in CeSb, especially since so much of the theory on these compounds hinges on the correct CEF splitting of the $\Gamma_7$ and $\Gamma_8$ manifolds of the $f$-states. 

\begin{figure*}[t]
\begin{center}
\includegraphics[width=16cm]{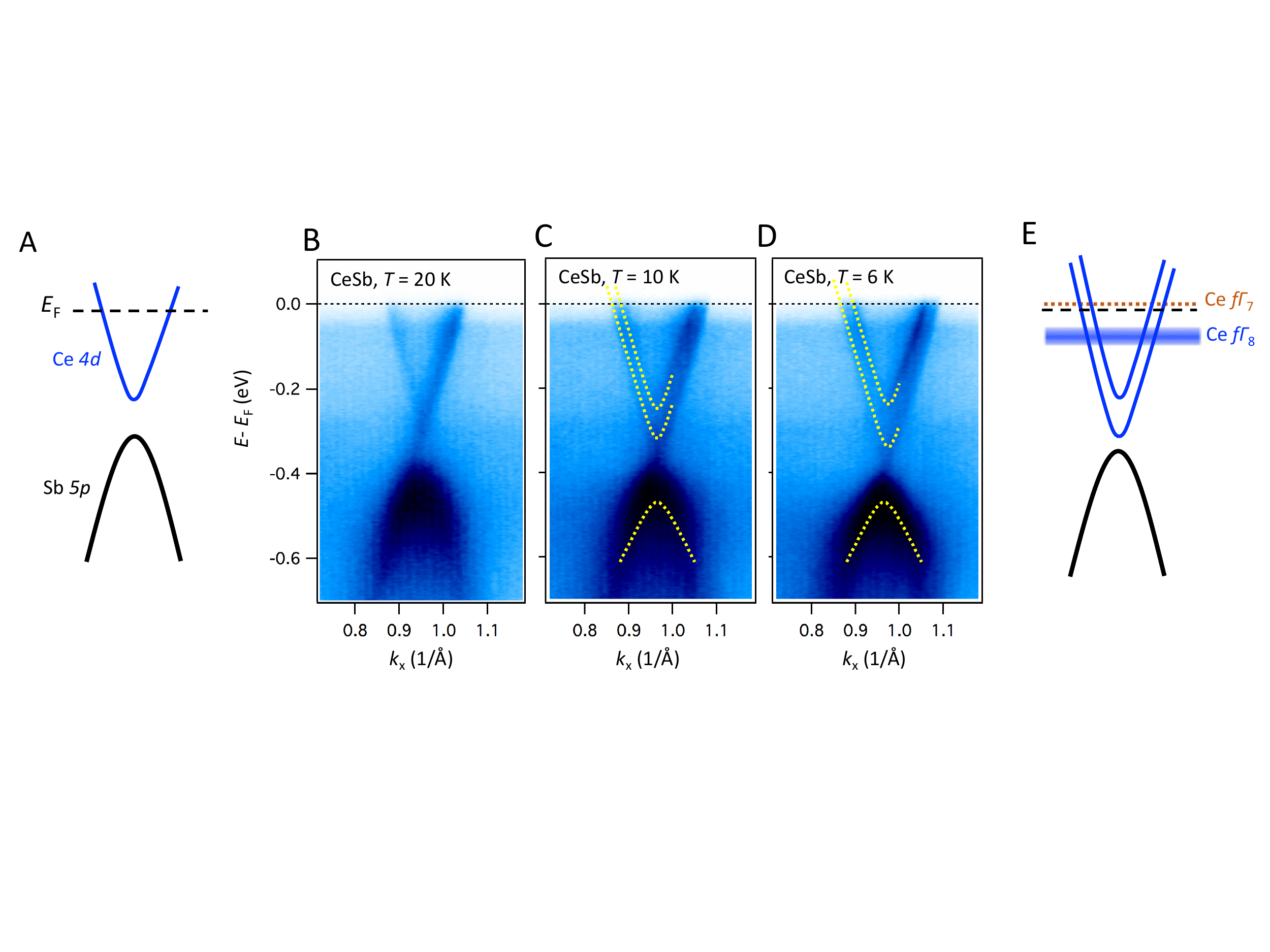}
\caption{{\bf Observation of magnetic exchange splitting at $X$-point}  Schematic of band structure, based on our ARPES data, on CeSb {\bf A} at the high temperature and {\bf E} at the low temperature near $X$-points. {\bf B to D} ARPES data taken at $hv$ = 88 eV near the $X$-point for the selected temperatures (indicated at the upper left). A clear signature of band splitting has been detected at $T$ = 6 K owing to Zeeman splitting, but disappeared above than $T^0_N$. 
}
\label{fig:fig4}
\end{center}
\end{figure*}

In Figure \ref{fig:fig4}, we show the evolution of the band structure near the $X$-point of the BZ. At the first transition $T_N^0$, the $d$-bands suddenly split, with a gap $\Delta$ that grows only slightly as the system is further cooled. Below $T_N^f$, where all PM layers have been extinguished, the splitting is $\sim 100$ meV, indicating a significant exchange coupling between the $f\Gamma_7$ ground state moments and the Ce $d$-electrons. Dynamical mean field theory (DMFT) calculations of a purely FM state are broadly consistent with the observed splitting (see Supplementary Information and schematic in Fig. \ref{fig:fig4}E). This suggests that the dominant effect on the band structure is the influence of the FM planes, and not the inter-layer antiferromagnetic (AFM) order which would tend to fold the zone. Importantly, there is no evidence of hybridization between the $f$-states and the $d$ electrons, which might be expected if Kondo-like behavior was active. Rather, the behavior of the $d$-electrons appear to be strongly exchange-coupled to the ordering of the $f\Gamma_7$ moments, consistent with non-Kondo magnetism~\cite{coqblin_exchange_1969,cooper_contrasting_1982,kioussis88}.

\begin{figure*}[t]
\begin{center}
\includegraphics[width=16cm]{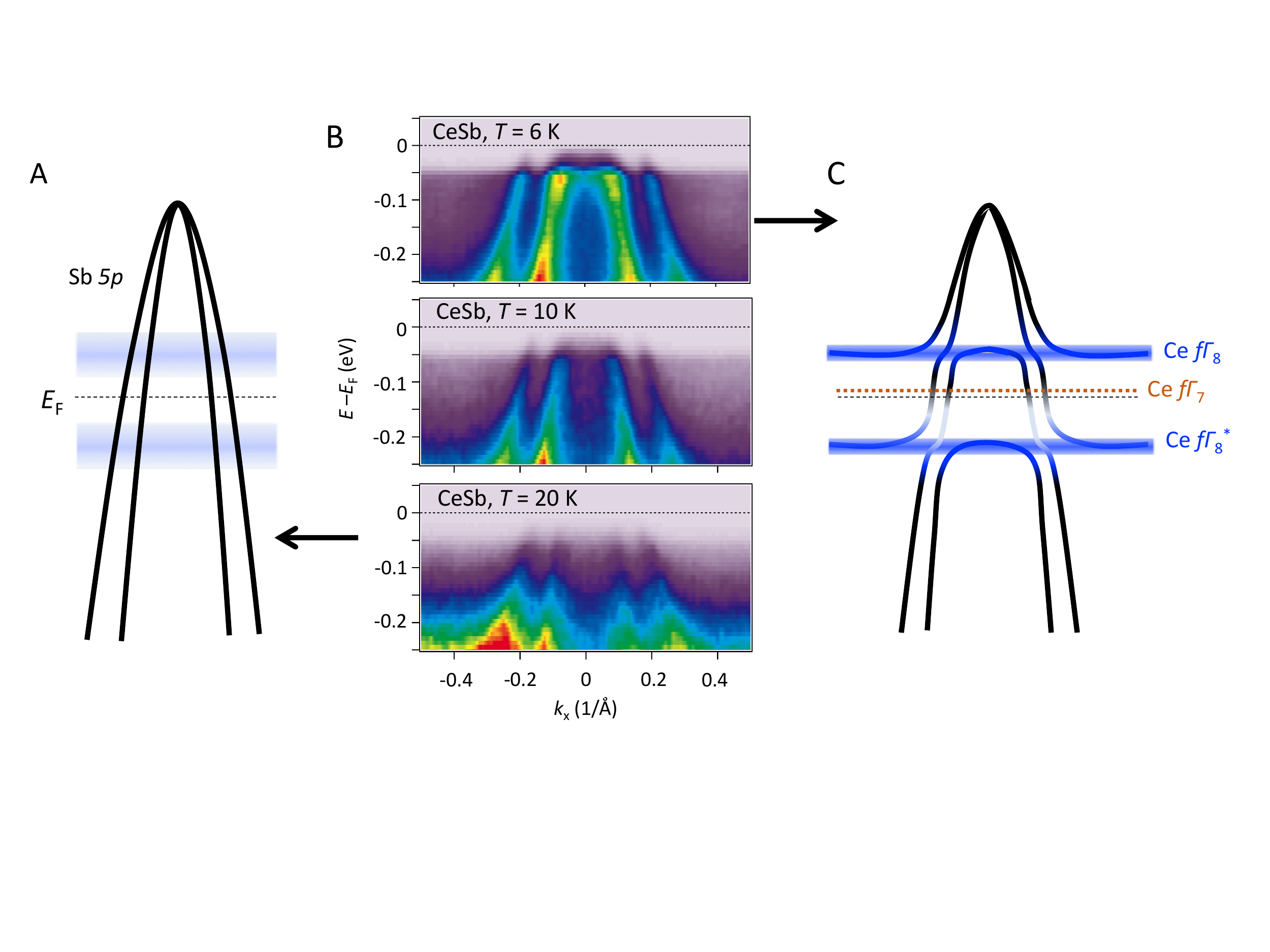}
\caption{{\bf Observation of $p-f$ hybridization at $\Gamma$-point} Schematics of the band structure at $\Gamma$-points, for CeSb {\bf A} at $T$ = 20 K and {\bf E} at $T$ = 6 K.  {\bf B to D} Experimental band structure of CeSb taken at selected temperatures marked at the upper right. The on-resonance photon energy is 122 eV for $k_Z$ at the high symmetry $\Gamma$-point of the bulk Brillouin zone (See Fig. S\ref{fig:fig3}A),  
}
\label{fig:fig3}
\end{center}
\end{figure*}

However, the behavior of the Sb $p$-bands at the $\Gamma$-point of the Brillouin zone stands in stark contrast to that of the Ce $d$-bands at $X$-point. Figure \ref{fig:fig3} illustrates the temperature dependence of the two hole-like bands as the temperature is lowered. At high temperatures $T>20 K$, the band structure and Fermi surface appears very similar to the non-magnetic analogue of this material LaSb (see Supplementary Information), with two $p$-bands crossing the Fermi energy. These form hole-like Fermi surfaces with no apparent $f$-character. As the temperature is lowered, the bands appear to curve at points where they cross the $f$-states, particularly the inner $p$-band with the $f\Gamma_8$ electrons. Such hybridization is consistent with the symmetry-enhanced interaction between the $f\Gamma_8$ states and the $p$-bands with $\Gamma_8$ character. Importantly we do not see the exchange splitting of the kind observed at $X$-point. Since we resonantly enhance the signal from the $f$-states, our data provides direct evidence that the dominant interaction between the $f$-states and the $p$-states is in the hybridization channel, as opposed to the exchange splitting seen at $X$. We are thus able to make an important statement about the physics of CeSb; the interactions between the $f$-electrons is strong with both $p$- and $d$-states, but they just choose different channels.

{\bf Discussion} Understanding the complex magnetism of CeSb depends on a complete picture of how the $f$-electrons interact with the electron and hole-like pockets~\cite{Takahashi85, kioussis88, kasuya93}. By symmetry, it is natural that the $p$-bands have a strong interaction with the $f\Gamma_8$ states,  just as we observe. However, given the large CEF indicated by our data, it seems less likely that this $p-f$ hybridization will drive admixing of (part of) the $f\Gamma_8$ excited states with the $f\Gamma_7$ ground state as required in some theoretical pictures of these materials~\cite{Takahashi85}. While there may be some magnetic interplay between the hybridized $p$-electrons and the exchange splitting of the $d$-electrons, the main result of this study is that the ground state $f$-moments interact in demonstrably different ways with the different carriers. The simultaneous activity of these interaction channels is quantum mechanically allowed because they occur in different parts of momentum space, painting an appealing picture for the CeSb's apparent duplicitous properties; coexisting signatures in the transport and magnetism that have previously been associated with Kondo-like screening and magnetic order, could be an indirect manifestation of coexisting paths of interaction.

A Kondo resonance peak at $E_F$ is however, not evident in the spectroscopy (Fig.~\ref{fig:fig2}A). This suggests that Kondo-like exchange must be weak, but not necessarily non-existent. The presence of such interactions may be indirectly evident by the presence final state effects that allow the observation of CEF and SO excited states, as described above. More interesting are the apparent momentum dependent interactions, whose mechanism may be analogous to processes in FM-Kondo materials, that have only recently been understood ~\cite{pruschke_low-energy_2000,liu13, pruschke12,golez13,si10}. In these materials, the Kondo interactions is spin-selective, allowing FM to coexist with partial Kondo screening. In CeSb, the interactions are orbitally-selective, weaving FM layers with PM layers and magnetic order with partial Kondo screening. 

Our observations in CeSb suggest an interesting paradigm for understanding the coexistence of many-body states broadly seen in correlated electron systems. A persistent question in the physics of heavy fermion materials has been why some itinerant quasiparticles are more susceptible to forming a Kondo coherence than others, which is evident in the coexistence of both heavy and light bands~\cite{fisk96}. Even though the Kondo signatures of CeSb are extremely weak, the material nevertheless provides a clue as to how this can occur; $f$-electrons can interact in different channels, depending on the orbital symmetry of the itinerant states themselves. The duplicitous nature of CeSb, marked by the coexistence in real space of different many-body phenomena, is allowed by the separation of different channels of interaction in momentum space.

\sectionmark{\bf FIG. 3. Band splitting }

\sectionmark{\bf FIG. 4. Observation of $p-f$ hybridization at $\Gamma$-point }

\subsection{Methods}

{\bf Experimental.}
Single crystals of CeSb were synthesized using tin flux.   Cerium (99.8\%), antimony (99.9999\%), and tin (99.999\%) (all from Alfa Aesar) were added to an alumina crucible in a molar ratio of 1:1:20.  The crucible was sealed in an evacuated quartz ampule before being heated over 8 hours to 1150 $^{\circ}$C, where it dwelled for 24 hours.  Next, the ampule was cooled to 800 $^{\circ}$C over 24 hours, and then it was centrifuged to remove excess tin.  This procedure yielded 5-10 mm single crystals.

	Temperature-dependent ARPES measurements were performed at the MERLIN beamline 4.0.3 of the Advanced Light Source (ALS) employing both linear horizontal and linear vertical polarization from an elliptically polarized undulator. A Scienta R8000 electron spectrometer with 2D parallel detection of electron kinetic energy and angle in combination with a six-axis helium cryostat goniometer with 6 K base temperature and $<5\times10^{-11}$ torr base pressure. Total energy resolution of approximately 15 meV was used for measurements at h$\nu$ = 122 eV corresponding to the Ce 4$d$-4$f$ resonant enhancement of the $f$ photoionization cross section.

\subsection*{Acknowledgements}
We would like to thank Paul Canfield, Joe Checkelsky and Piers Coleman for help discussions. This research is funded in part by the Gordon and Betty Moore Foundation's EPiQS Initiative through Grant No. GBMF4374 and U.S. DOE under contract No. DE-AC02-05-CH11231 to S.J., C.J., S.D., and J.G.A. R.K. acknowleges support by the National Science Foundation (NSF) Graduate Research Fellowship, Grant No. DGE-1106400. J.H.S. and J.H. was supported by the National Research Foundation of Korea (NRF) grant funded by the Korea government (MSIP) (No. 2015R1A2A1A15051540). The work at Rice University is supported by the NSF grant No. DMR-1611392, the Robert A. Welch Foundation Grant No. C-1411, the ARO Grant No. W911NF-14-1-0525, and a QuantEmX grant from ICAM and the Gordon and Betty Moore Foundation through Grant No. GBMF5305. Q.S. acknowledges the hospitality of University of California at Berkeley. 



\begin{thebibliography}{30}%
\makeatletter
\providecommand \@ifxundefined [1]{%
 \@ifx{#1\undefined}
}%
\providecommand \@ifnum [1]{%
 \ifnum #1\expandafter \@firstoftwo
 \else \expandafter \@secondoftwo
 \fi
}%
\providecommand \@ifx [1]{%
 \ifx #1\expandafter \@firstoftwo
 \else \expandafter \@secondoftwo
 \fi
}%
\providecommand \natexlab [1]{#1}%
\providecommand \enquote  [1]{``#1''}%
\providecommand \bibnamefont  [1]{#1}%
\providecommand \bibfnamefont [1]{#1}%
\providecommand \citenamefont [1]{#1}%
\providecommand \href@noop [0]{\@secondoftwo}%
\providecommand \href [0]{\begingroup \@sanitize@url \@href}%
\providecommand \@href[1]{\@@startlink{#1}\@@href}%
\providecommand \@@href[1]{\endgroup#1\@@endlink}%
\providecommand \@sanitize@url [0]{\catcode `\\12\catcode `\$12\catcode
  `\&12\catcode `\#12\catcode `\^12\catcode `\_12\catcode `\%12\relax}%
\providecommand \@@startlink[1]{}%
\providecommand \@@endlink[0]{}%
\providecommand \url  [0]{\begingroup\@sanitize@url \@url }%
\providecommand \@url [1]{\endgroup\@href {#1}{\urlprefix }}%
\providecommand \urlprefix  [0]{URL }%
\providecommand \Eprint [0]{\href }%
\providecommand \doibase [0]{http://dx.doi.org/}%
\providecommand \selectlanguage [0]{\@gobble}%
\providecommand \bibinfo  [0]{\@secondoftwo}%
\providecommand \bibfield  [0]{\@secondoftwo}%
\providecommand \translation [1]{[#1]}%
\providecommand \BibitemOpen [0]{}%
\providecommand \bibitemStop [0]{}%
\providecommand \bibitemNoStop [0]{.\EOS\space}%
\providecommand \EOS [0]{\spacefactor3000\relax}%
\providecommand \BibitemShut  [1]{\csname bibitem#1\endcsname}%
\let\auto@bib@innerbib\@empty
\bibitem [{\citenamefont {Kondo}(1964)}]{kondo64}%
  \BibitemOpen
  \bibfield  {author} {\bibinfo {author} {\bibfnamefont {J.}~\bibnamefont
  {Kondo}},\ }\href@noop {} {\bibfield  {journal} {\bibinfo  {journal}
  {Progress of Theoretical Physics}\ }\textbf {\bibinfo {volume} {32}},\
  \bibinfo {pages} {37} (\bibinfo {year} {1964})}\BibitemShut {NoStop}%
\bibitem [{\citenamefont {Jullien}\ \emph {et~al.}(1977)\citenamefont
  {Jullien}, \citenamefont {Fields},\ and\ \citenamefont
  {Doniach}}]{doniach77_1}%
  \BibitemOpen
  \bibfield  {author} {\bibinfo {author} {\bibfnamefont {R.}~\bibnamefont
  {Jullien}}, \bibinfo {author} {\bibfnamefont {J.~N.}\ \bibnamefont {Fields}},
  \ and\ \bibinfo {author} {\bibfnamefont {S.}~\bibnamefont {Doniach}},\
  }\href@noop {} {\bibfield  {journal} {\bibinfo  {journal} {Phys. Rev. Lett.}\
  }\textbf {\bibinfo {volume} {38}},\ \bibinfo {pages} {1500} (\bibinfo {year}
  {1977})}\BibitemShut {NoStop}%
\bibitem [{\citenamefont {Mott}(1974)}]{mott74}%
  \BibitemOpen
  \bibfield  {author} {\bibinfo {author} {\bibfnamefont {N.~F.}\ \bibnamefont
  {Mott}},\ }\href@noop {} {\bibfield  {journal} {\bibinfo  {journal} {Phil.
  Mag.}\ }\textbf {\bibinfo {volume} {30}},\ \bibinfo {pages} {403} (\bibinfo
  {year} {1974})}\BibitemShut {NoStop}%
\bibitem [{\citenamefont {Doniach}(1977)}]{doniach77}%
  \BibitemOpen
  \bibfield  {author} {\bibinfo {author} {\bibfnamefont {S.}~\bibnamefont
  {Doniach}},\ }\href@noop {} {\bibfield  {journal} {\bibinfo  {journal}
  {Physica B}\ }\textbf {\bibinfo {volume} {91}},\ \bibinfo {pages} {231}
  (\bibinfo {year} {1977})}\BibitemShut {NoStop}%
\bibitem [{\citenamefont {Coqblin}\ and\ \citenamefont
  {Schrieffer}(1969)}]{coqblin_exchange_1969}%
  \BibitemOpen
  \bibfield  {author} {\bibinfo {author} {\bibfnamefont {B.}~\bibnamefont
  {Coqblin}}\ and\ \bibinfo {author} {\bibfnamefont {J.~R.}\ \bibnamefont
  {Schrieffer}},\ }\href {\doibase 10.1103/PhysRev.185.847} {\bibfield
  {journal} {\bibinfo  {journal} {Physical Review}\ }\textbf {\bibinfo {volume}
  {185}},\ \bibinfo {pages} {847} (\bibinfo {year} {1969})}\BibitemShut
  {NoStop}%
\bibitem [{\citenamefont {Cooper}(1982)}]{cooper_contrasting_1982}%
  \BibitemOpen
  \bibfield  {author} {\bibinfo {author} {\bibfnamefont {B.~R.}\ \bibnamefont
  {Cooper}},\ }\href {\doibase 10.1016/0304-8853(82)90245-1} {\bibfield
  {journal} {\bibinfo  {journal} {Journal of Magnetism and Magnetic Materials}\
  }\textbf {\bibinfo {volume} {29}},\ \bibinfo {pages} {230} (\bibinfo {year}
  {1982})}\BibitemShut {NoStop}%
\bibitem [{\citenamefont {Kioussis}\ \emph {et~al.}(1998)\citenamefont
  {Kioussis}, \citenamefont {Cooper},\ and\ \citenamefont
  {Banerjea}}]{kioussis88}%
  \BibitemOpen
  \bibfield  {author} {\bibinfo {author} {\bibfnamefont {N.}~\bibnamefont
  {Kioussis}}, \bibinfo {author} {\bibfnamefont {B.~R.}\ \bibnamefont
  {Cooper}}, \ and\ \bibinfo {author} {\bibfnamefont {A.}~\bibnamefont
  {Banerjea}},\ }\href@noop {} {\bibfield  {journal} {\bibinfo  {journal}
  {Phys. Rev. B}\ }\textbf {\bibinfo {volume} {38}},\ \bibinfo {pages} {9132}
  (\bibinfo {year} {1998})}\BibitemShut {NoStop}%
\bibitem [{\citenamefont {Guo}\ \emph {et~al.}(2017)\citenamefont {Guo},
  \citenamefont {Cao}, \citenamefont {Smidman}, \citenamefont {Wu},
  \citenamefont {Zhang}, \citenamefont {Steglich}, \citenamefont {Zhang},\ and\
  \citenamefont {Yuan}}]{guo_possible_2017}%
  \BibitemOpen
  \bibfield  {author} {\bibinfo {author} {\bibfnamefont {C.}~\bibnamefont
  {Guo}}, \bibinfo {author} {\bibfnamefont {C.}~\bibnamefont {Cao}}, \bibinfo
  {author} {\bibfnamefont {M.}~\bibnamefont {Smidman}}, \bibinfo {author}
  {\bibfnamefont {F.}~\bibnamefont {Wu}}, \bibinfo {author} {\bibfnamefont
  {Y.}~\bibnamefont {Zhang}}, \bibinfo {author} {\bibfnamefont
  {F.}~\bibnamefont {Steglich}}, \bibinfo {author} {\bibfnamefont {F.-C.}\
  \bibnamefont {Zhang}}, \ and\ \bibinfo {author} {\bibfnamefont
  {H.}~\bibnamefont {Yuan}},\ }\href {\doibase 10.1038/s41535-017-0038-3}
  {\bibfield  {journal} {\bibinfo  {journal} {npj Quantum Materials}\ }\textbf
  {\bibinfo {volume} {2}},\ \bibinfo {pages} {39} (\bibinfo {year}
  {2017})}\BibitemShut {NoStop}%
\bibitem [{\citenamefont {Rossat-Mignod}\ \emph {et~al.}(1977)\citenamefont
  {Rossat-Mignod}, \citenamefont {Burlet}, \citenamefont {Villain},
  \citenamefont {Bartholin}, \citenamefont {Tcheng-Si}, \citenamefont
  {Florence},\ and\ \citenamefont {Vogt}}]{Rossat77}%
  \BibitemOpen
  \bibfield  {author} {\bibinfo {author} {\bibfnamefont {J.}~\bibnamefont
  {Rossat-Mignod}}, \bibinfo {author} {\bibfnamefont {P.}~\bibnamefont
  {Burlet}}, \bibinfo {author} {\bibfnamefont {J.}~\bibnamefont {Villain}},
  \bibinfo {author} {\bibfnamefont {H.}~\bibnamefont {Bartholin}}, \bibinfo
  {author} {\bibfnamefont {W.}~\bibnamefont {Tcheng-Si}}, \bibinfo {author}
  {\bibfnamefont {D.}~\bibnamefont {Florence}}, \ and\ \bibinfo {author}
  {\bibfnamefont {O.}~\bibnamefont {Vogt}},\ }\href@noop {} {\bibfield
  {journal} {\bibinfo  {journal} {Phys. Rev. B}\ }\textbf {\bibinfo {volume}
  {16}},\ \bibinfo {pages} {440} (\bibinfo {year} {1977})}\BibitemShut
  {NoStop}%
\bibitem [{\citenamefont {Wiener}\ \emph {et~al.}(2000)\citenamefont {Wiener},
  \ and\ \citenamefont
  {Canfield}}]{canfield00}%
  \BibitemOpen
  \bibfield  {author} {\bibinfo {author} {\bibfnamefont {T.A.}~\bibnamefont
  {Wiener}}, \bibinfo {author} {\bibfnamefont {P.C.}~\bibnamefont {Canfield}}, \ }\href@noop {}
  {\bibfield  {journal} {\bibinfo  {journal} {J. Alloys Compd.}\ }\textbf {\bibinfo {volume}
  {303-304}},\ \bibinfo {pages} {505} (\bibinfo
  {year} {2000})}\BibitemShut {NoStop}%
\bibitem [{\citenamefont {Ye}\ \emph {et~al.}(2017)\citenamefont {Ye},
  \citenamefont {Suzuki}, \citenamefont {Wicker},\ and\ \citenamefont
  {Checkelsky}}]{chechelsky17}%
  \BibitemOpen
  \bibfield  {author} {\bibinfo {author} {\bibfnamefont {L.}~\bibnamefont
  {Ye}}, \bibinfo {author} {\bibfnamefont {T.}~\bibnamefont {Suzuki}}, \bibinfo
  {author} {\bibfnamefont {C.~R.}\ \bibnamefont {Wicker}}, \ and\ \bibinfo
  {author} {\bibfnamefont {J.~G.}\ \bibnamefont {Checkelsky}},\ }\href@noop {}
  {\bibfield  {journal} {\bibinfo  {journal} {arXiv: 1704.04226v1}\ } (\bibinfo
  {year} {2017})}\BibitemShut {NoStop}%
\bibitem [{\citenamefont {Hundley}\ \emph {et~al.}(2004)\citenamefont
  {Hundley}, \citenamefont {Malinowski}, \citenamefont {Pagliuso},
  \citenamefont {Sarrao},\ and\ \citenamefont {Thompson}}]{hundley04}%
  \BibitemOpen
  \bibfield  {author} {\bibinfo {author} {\bibfnamefont {M.~F.}\ \bibnamefont
  {Hundley}}, \bibinfo {author} {\bibfnamefont {A.}~\bibnamefont {Malinowski}},
  \bibinfo {author} {\bibfnamefont {P.~G.}\ \bibnamefont {Pagliuso}}, \bibinfo
  {author} {\bibfnamefont {J.~L.}\ \bibnamefont {Sarrao}}, \ and\ \bibinfo
  {author} {\bibfnamefont {J.~D.}\ \bibnamefont {Thompson}},\ }\href@noop {}
  {\bibfield  {journal} {\bibinfo  {journal} {Phys. Rev. B}\ }\textbf {\bibinfo
  {volume} {70}},\ \bibinfo {pages} {035113} (\bibinfo {year}
  {2004})}\BibitemShut {NoStop}%
\bibitem [{\citenamefont {K.Yamamoto}\ and\ \citenamefont
  {Ueda}(1990)}]{ueda90}%
  \BibitemOpen
  \bibfield  {author} {\bibinfo {author} {\bibnamefont {K.Yamamoto}}\ and\
  \bibinfo {author} {\bibfnamefont {K.}~\bibnamefont {Ueda}},\ }\href@noop {}
  {\bibfield  {journal} {\bibinfo  {journal} {J. Phys. Soc. Jpn.}\ }\textbf
  {\bibinfo {volume} {59}},\ \bibinfo {pages} {3284} (\bibinfo {year}
  {1990})}\BibitemShut {NoStop}%
\bibitem [{\citenamefont {Heer}\ \emph {et~al.}(1979)\citenamefont {Heer},
  \citenamefont {Furrer}, \citenamefont {Halg},\ and\ \citenamefont
  {Vogt}}]{heer_neutron_1979}%
  \BibitemOpen
  \bibfield  {author} {\bibinfo {author} {\bibfnamefont {H.}~\bibnamefont
  {Heer}}, \bibinfo {author} {\bibfnamefont {A.}~\bibnamefont {Furrer}},
  \bibinfo {author} {\bibfnamefont {W.}~\bibnamefont {Halg}}, \ and\ \bibinfo
  {author} {\bibfnamefont {O.}~\bibnamefont {Vogt}},\ }\href {\doibase
  10.1088/0022-3719/12/23/025} {\bibfield  {journal} {\bibinfo  {journal}
  {Journal of Physics C: Solid State Physics}\ }\textbf {\bibinfo {volume}
  {12}},\ \bibinfo {pages} {5207} (\bibinfo {year} {1979})}\BibitemShut
  {NoStop}%
\bibitem [{\citenamefont {Kasaya}\ \emph {et~al.}(1983)\citenamefont {Kasaya},
  \citenamefont {Iga}, \citenamefont {Negishi}, \citenamefont {Nakai},\ and\
  \citenamefont {Kasuya}}]{kasaya83}%
  \BibitemOpen
  \bibfield  {author} {\bibinfo {author} {\bibfnamefont {M.}~\bibnamefont
  {Kasaya}}, \bibinfo {author} {\bibfnamefont {F.}~\bibnamefont {Iga}},
  \bibinfo {author} {\bibfnamefont {K.}~\bibnamefont {Negishi}}, \bibinfo
  {author} {\bibfnamefont {S.}~\bibnamefont {Nakai}}, \ and\ \bibinfo {author}
  {\bibfnamefont {T.}~\bibnamefont {Kasuya}},\ }\href@noop {} {\bibfield
  {journal} {\bibinfo  {journal} {J. Magn Magn, Mater.}\ }\textbf {\bibinfo
  {volume} {31-34}},\ \bibinfo {pages} {437} (\bibinfo {year}
  {1983})}\BibitemShut {NoStop}%
\bibitem [{\citenamefont {Takahashi}\ and\ \citenamefont
  {Kasuya}(1985)}]{Takahashi85}%
  \BibitemOpen
  \bibfield  {author} {\bibinfo {author} {\bibfnamefont {H.}~\bibnamefont
  {Takahashi}}\ and\ \bibinfo {author} {\bibfnamefont {T.}~\bibnamefont
  {Kasuya}},\ }\href@noop {} {\bibfield  {journal} {\bibinfo  {journal} {J.
  Phys. C}\ }\textbf {\bibinfo {volume} {13}},\ \bibinfo {pages} {6147}
  (\bibinfo {year} {1985})}\BibitemShut {NoStop}%
\bibitem [{\citenamefont {Alidoust}\ \emph {et~al.}(2016)\citenamefont
  {Alidoust}, \citenamefont {Alexandradinata}, \citenamefont {Xu},
  \citenamefont {Belopolski}, \citenamefont {Kushwaha}, \citenamefont {Zeng},
  \citenamefont {Neupane}, \citenamefont {Bian}, \citenamefont {Liu},
  \citenamefont {Sanchez}, \citenamefont {Shibayev}, \citenamefont {Zheng},
  \citenamefont {Fu}, \citenamefont {Lin}, \citenamefont {Cava},\ and\
  \citenamefont {Hasan}}]{hasan16}%
  \BibitemOpen
  \bibfield  {author} {\bibinfo {author} {\bibfnamefont {N.}~\bibnamefont
  {Alidoust}}, \bibinfo {author} {\bibfnamefont {A.}~\bibnamefont
  {Alexandradinata}}, \bibinfo {author} {\bibfnamefont {S.-Y.}\ \bibnamefont
  {Xu}}, \bibinfo {author} {\bibfnamefont {I.}~\bibnamefont {Belopolski}},
  \bibinfo {author} {\bibfnamefont {S.~K.}\ \bibnamefont {Kushwaha}}, \bibinfo
  {author} {\bibfnamefont {M.}~\bibnamefont {Zeng}}, \bibinfo {author}
  {\bibfnamefont {M.}~\bibnamefont {Neupane}}, \bibinfo {author} {\bibfnamefont
  {G.}~\bibnamefont {Bian}}, \bibinfo {author} {\bibfnamefont {C.}~\bibnamefont
  {Liu}}, \bibinfo {author} {\bibfnamefont {D.~S.}\ \bibnamefont {Sanchez}},
  \bibinfo {author} {\bibfnamefont {P.~P.}\ \bibnamefont {Shibayev}}, \bibinfo
  {author} {\bibfnamefont {H.}~\bibnamefont {Zheng}}, \bibinfo {author}
  {\bibfnamefont {L.}~\bibnamefont {Fu}}, \bibinfo {author} {\bibfnamefont
  {A.~B.~H.}\ \bibnamefont {Lin}}, \bibinfo {author} {\bibfnamefont {R.~J.}\
  \bibnamefont {Cava}}, \ and\ \bibinfo {author} {\bibfnamefont {M.~Z.}\
  \bibnamefont {Hasan}},\ }\href@noop {} {\bibfield  {journal} {\bibinfo
  {journal} {arXiv:1604.08571v1}\ } (\bibinfo {year} {2016})}\BibitemShut
  {NoStop}%
\bibitem [{\citenamefont {Wu}\ \emph {et~al.}(2017)\citenamefont {Wu},
  \citenamefont {Lee}, \citenamefont {Kong}, \citenamefont {Mou}, \citenamefont
  {Jiang}, \citenamefont {Huang}, \citenamefont {Bud'ko}, \citenamefont
  {Canfield},\ and\ \citenamefont {Kaminski}}]{kaminski17}%
  \BibitemOpen
  \bibfield  {author} {\bibinfo {author} {\bibfnamefont {Y.}~\bibnamefont
  {Wu}}, \bibinfo {author} {\bibfnamefont {Y.}~\bibnamefont {Lee}}, \bibinfo
  {author} {\bibfnamefont {T.}~\bibnamefont {Kong}}, \bibinfo {author}
  {\bibfnamefont {D.}~\bibnamefont {Mou}}, \bibinfo {author} {\bibfnamefont
  {R.}~\bibnamefont {Jiang}}, \bibinfo {author} {\bibfnamefont
  {L.}~\bibnamefont {Huang}}, \bibinfo {author} {\bibfnamefont {S.~L.}\
  \bibnamefont {Bud'ko}}, \bibinfo {author} {\bibfnamefont {P.~C.}\
  \bibnamefont {Canfield}}, \ and\ \bibinfo {author} {\bibfnamefont
  {A.}~\bibnamefont {Kaminski}},\ }\href@noop {} {\bibfield  {journal}
  {\bibinfo  {journal} {arXiv:1704.06237v1}\ } (\bibinfo {year}
  {2017})}\BibitemShut {NoStop}%
\bibitem [{\citenamefont {Oinuma}\ \emph {et~al.}(2017)\citenamefont {Oinuma},
  \citenamefont {Souma}, \citenamefont {Takane}, \citenamefont {Nakamura},
  \citenamefont {Nakayama}, \citenamefont {Mitsuhashi}, \citenamefont {Horiba},
  \citenamefont {Kumigashira}, \citenamefont {Yoshida}, \citenamefont {Ochiai},
  \citenamefont {Takahashi},\ and\ \citenamefont {Sato}}]{sato17}%
  \BibitemOpen
  \bibfield  {author} {\bibinfo {author} {\bibfnamefont {H.}~\bibnamefont
  {Oinuma}}, \bibinfo {author} {\bibfnamefont {S.}~\bibnamefont {Souma}},
  \bibinfo {author} {\bibfnamefont {D.}~\bibnamefont {Takane}}, \bibinfo
  {author} {\bibfnamefont {T.}~\bibnamefont {Nakamura}}, \bibinfo {author}
  {\bibfnamefont {K.}~\bibnamefont {Nakayama}}, \bibinfo {author}
  {\bibfnamefont {T.}~\bibnamefont {Mitsuhashi}}, \bibinfo {author}
  {\bibfnamefont {K.}~\bibnamefont {Horiba}}, \bibinfo {author} {\bibfnamefont
  {H.}~\bibnamefont {Kumigashira}}, \bibinfo {author} {\bibfnamefont
  {M.}~\bibnamefont {Yoshida}}, \bibinfo {author} {\bibfnamefont
  {A.}~\bibnamefont {Ochiai}}, \bibinfo {author} {\bibfnamefont
  {T.}~\bibnamefont {Takahashi}}, \ and\ \bibinfo {author} {\bibfnamefont
  {T.}~\bibnamefont {Sato}},\ }\href@noop {} {\bibfield  {journal} {\bibinfo
  {journal} {arXiv: 1707.05100v1}\ } (\bibinfo {year} {2017})}\BibitemShut
  {NoStop}%
\bibitem [{\citenamefont {Takayama}\ \emph {et~al.}(2009)\citenamefont
  {Takayama}, \citenamefont {Souma}, \citenamefont {Sato}, \citenamefont
  {Arakane},\ and\ \citenamefont {Takahashi}}]{takahashi09}%
  \BibitemOpen
  \bibfield  {author} {\bibinfo {author} {\bibfnamefont {A.}~\bibnamefont
  {Takayama}}, \bibinfo {author} {\bibfnamefont {S.}~\bibnamefont {Souma}},
  \bibinfo {author} {\bibfnamefont {T.}~\bibnamefont {Sato}}, \bibinfo {author}
  {\bibfnamefont {T.}~\bibnamefont {Arakane}}, \ and\ \bibinfo {author}
  {\bibfnamefont {T.}~\bibnamefont {Takahashi}},\ }\href@noop {} {\bibfield
  {journal} {\bibinfo  {journal} {J. Phys. Soc. Jpn.}\ }\textbf {\bibinfo
  {volume} {78}},\ \bibinfo {pages} {073702} (\bibinfo {year}
  {2009})}\BibitemShut {NoStop}%
\bibitem [{\citenamefont {H$\ddot{\mathrm{u}}$fner}(1995)}]{hufner95}%
  \BibitemOpen
  \bibfield  {author} {\bibinfo {author} {\bibfnamefont {S.}~\bibnamefont
  {H$\ddot{\mathrm{u}}$fner}},\ }\href@noop {} {\emph {\bibinfo {title}
  {Photoelectron Spectroscopy}}}\ (\bibinfo  {publisher} {Springer, Berlin},\
  \bibinfo {year} {1995})\BibitemShut {NoStop}%
\bibitem [{\citenamefont {Rossat-Mignod}\ \emph {et~al.}(1985)\citenamefont
  {Rossat-Mignod}, \citenamefont {Effantin}, \citenamefont {Burlet},
  \citenamefont {Chattopadhyay}, \citenamefont {Regnault}, \citenamefont
  {Bartholin}, \citenamefont {Vettier}, \citenamefont {Vogt}, \citenamefont
  {Ravot},\ and\ \citenamefont {Achard}}]{Rossat85}%
  \BibitemOpen
  \bibfield  {author} {\bibinfo {author} {\bibfnamefont {J.}~\bibnamefont
  {Rossat-Mignod}}, \bibinfo {author} {\bibfnamefont {J.~M.}\ \bibnamefont
  {Effantin}}, \bibinfo {author} {\bibfnamefont {P.}~\bibnamefont {Burlet}},
  \bibinfo {author} {\bibfnamefont {T.}~\bibnamefont {Chattopadhyay}}, \bibinfo
  {author} {\bibfnamefont {L.~P.}\ \bibnamefont {Regnault}}, \bibinfo {author}
  {\bibfnamefont {H.}~\bibnamefont {Bartholin}}, \bibinfo {author}
  {\bibfnamefont {C.}~\bibnamefont {Vettier}}, \bibinfo {author} {\bibfnamefont
  {O.}~\bibnamefont {Vogt}}, \bibinfo {author} {\bibfnamefont {D.}~\bibnamefont
  {Ravot}}, \ and\ \bibinfo {author} {\bibfnamefont {J.~C.}\ \bibnamefont
  {Achard}},\ }\href@noop {} {\bibfield  {journal} {\bibinfo  {journal} {J.
  Magn. Magn. Mater.}\ }\textbf {\bibinfo {volume} {52}},\ \bibinfo {pages}
  {111} (\bibinfo {year} {1985})}\BibitemShut {NoStop}%
\bibitem [{\citenamefont {Koitzsch}\ \emph {et~al.}(2016)\citenamefont
  {Koitzsch}, \citenamefont {Heming}, \citenamefont {Knupfer}, \citenamefont
  {Buchner}, \citenamefont {Portnichenko}, \citenamefont {Dukhnenko},
  \citenamefont {Shitsevalova}, \citenamefont {Filipov}, \citenamefont {Lev},
  \citenamefont {Strocov}, \citenamefont {Ollivier},\ and\ \citenamefont
  {Inosov}}]{inosov16}%
  \BibitemOpen
  \bibfield  {author} {\bibinfo {author} {\bibfnamefont {A.}~\bibnamefont
  {Koitzsch}}, \bibinfo {author} {\bibfnamefont {N.}~\bibnamefont {Heming}},
  \bibinfo {author} {\bibfnamefont {M.}~\bibnamefont {Knupfer}}, \bibinfo
  {author} {\bibfnamefont {B.}~\bibnamefont {Buchner}}, \bibinfo {author}
  {\bibfnamefont {P.}~\bibnamefont {Portnichenko}}, \bibinfo {author}
  {\bibfnamefont {A.}~\bibnamefont {Dukhnenko}}, \bibinfo {author}
  {\bibfnamefont {N.}~\bibnamefont {Shitsevalova}}, \bibinfo {author}
  {\bibfnamefont {V.}~\bibnamefont {Filipov}}, \bibinfo {author} {\bibfnamefont
  {L.}~\bibnamefont {Lev}}, \bibinfo {author} {\bibfnamefont {V.}~\bibnamefont
  {Strocov}}, \bibinfo {author} {\bibfnamefont {J.}~\bibnamefont {Ollivier}}, \
  and\ \bibinfo {author} {\bibfnamefont {D.}~\bibnamefont {Inosov}},\
  }\href@noop {} {\bibfield  {journal} {\bibinfo  {journal} {Nat. Comms.}\
  }\textbf {\bibinfo {volume} {7}},\ \bibinfo {pages} {10876} (\bibinfo {year}
  {2016})}\BibitemShut {NoStop}%
\bibitem [{\citenamefont {Birgeneau}\ \emph {et~al.}(1973)\citenamefont
  {Birgeneau}, \citenamefont {Bucher}, \citenamefont {Maita}, \citenamefont
  {Passell},\ and\ \citenamefont {Turberfield}}]{birgeneau_crystal_1973}%
  \BibitemOpen
  \bibfield  {author} {\bibinfo {author} {\bibfnamefont {R.~J.}\ \bibnamefont
  {Birgeneau}}, \bibinfo {author} {\bibfnamefont {E.}~\bibnamefont {Bucher}},
  \bibinfo {author} {\bibfnamefont {J.~P.}\ \bibnamefont {Maita}}, \bibinfo
  {author} {\bibfnamefont {L.}~\bibnamefont {Passell}}, \ and\ \bibinfo
  {author} {\bibfnamefont {K.~C.}\ \bibnamefont {Turberfield}},\ }\href
  {\doibase 10.1103/PhysRevB.8.5345} {\bibfield  {journal} {\bibinfo  {journal}
  {Physical Review B}\ }\textbf {\bibinfo {volume} {8}},\ \bibinfo {pages}
  {5345} (\bibinfo {year} {1973})}\BibitemShut {NoStop}%
\bibitem [{\citenamefont {Kasuya}\ \emph {et~al.}(1993)\citenamefont {Kasuya},
  \citenamefont {Hasa}, \citenamefont {Kwon},\ and\ \citenamefont
  {Suzuki}}]{kasuya93}%
  \BibitemOpen
  \bibfield  {author} {\bibinfo {author} {\bibfnamefont {T.}~\bibnamefont
  {Kasuya}}, \bibinfo {author} {\bibfnamefont {Y.}~\bibnamefont {Hasa}},
  \bibinfo {author} {\bibfnamefont {Y.}~\bibnamefont {Kwon}}, \ and\ \bibinfo
  {author} {\bibfnamefont {T.}~\bibnamefont {Suzuki}},\ }\href@noop {}
  {\bibfield  {journal} {\bibinfo  {journal} {Physica B}\ }\textbf {\bibinfo
  {volume} {186-188}},\ \bibinfo {pages} {9} (\bibinfo {year}
  {1993})}\BibitemShut {NoStop}%
\bibitem [{\citenamefont {Pruschke}\ \emph {et~al.}(2000)\citenamefont
  {Pruschke}, \citenamefont {Bulla},\ and\ \citenamefont
  {Jarrell}}]{pruschke_low-energy_2000}%
  \BibitemOpen
  \bibfield  {author} {\bibinfo {author} {\bibfnamefont {T.}~\bibnamefont
  {Pruschke}}, \bibinfo {author} {\bibfnamefont {R.}~\bibnamefont {Bulla}}, \
  and\ \bibinfo {author} {\bibfnamefont {M.}~\bibnamefont {Jarrell}},\ }\href
  {\doibase 10.1103/PhysRevB.61.12799} {\bibfield  {journal} {\bibinfo
  {journal} {Physical Review B}\ }\textbf {\bibinfo {volume} {61}},\ \bibinfo
  {pages} {12799} (\bibinfo {year} {2000})}\BibitemShut {NoStop}%
\bibitem [{\citenamefont {Liu}\ \emph {et~al.}(2013)\citenamefont {Liu},
  \citenamefont {Zhang},\ and\ \citenamefont {Yu}}]{liu13}%
  \BibitemOpen
  \bibfield  {author} {\bibinfo {author} {\bibfnamefont {Y.}~\bibnamefont
  {Liu}}, \bibinfo {author} {\bibfnamefont {G.-M.}\ \bibnamefont {Zhang}}, \
  and\ \bibinfo {author} {\bibfnamefont {L.}~\bibnamefont {Yu}},\ }\href@noop
  {} {\bibfield  {journal} {\bibinfo  {journal} {Phys. Rev. B}\ }\textbf
  {\bibinfo {volume} {87}},\ \bibinfo {pages} {134409} (\bibinfo {year}
  {2013})}\BibitemShut {NoStop}%
\bibitem [{\citenamefont {Peters}\ \emph {et~al.}(2012)\citenamefont {Peters},
  \citenamefont {Kawakami},\ and\ \citenamefont {Pruschke}}]{pruschke12}%
  \BibitemOpen
  \bibfield  {author} {\bibinfo {author} {\bibfnamefont {R.}~\bibnamefont
  {Peters}}, \bibinfo {author} {\bibfnamefont {N.}~\bibnamefont {Kawakami}}, \
  and\ \bibinfo {author} {\bibfnamefont {T.}~\bibnamefont {Pruschke}},\
  }\href@noop {} {\bibfield  {journal} {\bibinfo  {journal} {Phys. Rev. Lett.}\
  }\textbf {\bibinfo {volume} {108}},\ \bibinfo {pages} {086402} (\bibinfo
  {year} {2012})}\BibitemShut {NoStop}%
\bibitem [{\citenamefont {Gole$\breve{z}$}\ and\ \citenamefont
  {$\breve{Z}$itko}(2013)}]{golez13}%
  \BibitemOpen
  \bibfield  {author} {\bibinfo {author} {\bibfnamefont {D.}~\bibnamefont
  {Gole$\breve{z}$}}\ and\ \bibinfo {author} {\bibfnamefont {R.}~\bibnamefont
  {$\breve{Z}$itko}},\ }\href@noop {} {\bibfield  {journal} {\bibinfo
  {journal} {Phys. Rev. B}\ }\textbf {\bibinfo {volume} {88}},\ \bibinfo
  {pages} {054431} (\bibinfo {year} {2013})}\BibitemShut {NoStop}%
\bibitem [{\citenamefont {Yamamoto}\ and\ \citenamefont {Si}(2010)}]{si10}%
  \BibitemOpen
  \bibfield  {author} {\bibinfo {author} {\bibfnamefont {S.~J.}\ \bibnamefont
  {Yamamoto}}\ and\ \bibinfo {author} {\bibfnamefont {Q.}~\bibnamefont {Si}},\
  }\href@noop {} {\bibfield  {journal} {\bibinfo  {journal} {Proc. Natl. Acad.
  Sci.}\ }\textbf {\bibinfo {volume} {107}},\ \bibinfo {pages} {15704}
  (\bibinfo {year} {2010})}\BibitemShut {NoStop}%
\bibitem [{\citenamefont {Movshovich}\ \emph {et~al.}(1996)\citenamefont
  {Movshovich}, \citenamefont {Graf}, \citenamefont {Mandrus}, \citenamefont
  {Thompson}, \citenamefont {Smith},\ and\ \citenamefont {Fisk}}]{fisk96}%
  \BibitemOpen
  \bibfield  {author} {\bibinfo {author} {\bibfnamefont {R.}~\bibnamefont
  {Movshovich}}, \bibinfo {author} {\bibfnamefont {T.}~\bibnamefont {Graf}},
  \bibinfo {author} {\bibfnamefont {D.}~\bibnamefont {Mandrus}}, \bibinfo
  {author} {\bibfnamefont {J.~D.}\ \bibnamefont {Thompson}}, \bibinfo {author}
  {\bibfnamefont {J.~L.}\ \bibnamefont {Smith}}, \ and\ \bibinfo {author}
  {\bibfnamefont {Z.}~\bibnamefont {Fisk}},\ }\href {\doibase
  10.1103/PhysRevB.53.8241} {\bibfield  {journal} {\bibinfo  {journal} {Phys.
  Rev. B}\ }\textbf {\bibinfo {volume} {53}},\ \bibinfo {pages} {8241}
  (\bibinfo {year} {1996})}\BibitemShut {NoStop}%
\end{thebibliography}
%

\end{document}